\documentclass[12pt]{article}      

\usepackage{amsmath}
\usepackage{epsfig}

\usepackage{graphicx}

\title{Kaleidoscope of Classical Images and Quantum Coherent States}

\author{ Oktay K Pashaev and  Ayg\"{u}l Ko\c{c}ak \\Department of Mathematics, Izmir Institute of Technology \\ Urla-Izmir, 35430, Turkey}

\begin{document}
\newcommand{\be}{\begin{equation}}
\newcommand{\ee}{\end{equation}}
\newcommand{\bea}{\begin{eqnarray}}
\newcommand{\eea}{\end{eqnarray}}
\newcommand{\disp}{\displaystyle}
\newcommand{\la}{\langle}
\newcommand{\ra}{\rangle}

\newtheorem{thm}{Theorem}[subsection]
\newtheorem{cor}[thm]{Corollary}
\newtheorem{lem}[thm]{Lemma}
\newtheorem{prop}[thm]{Proposition}
\newtheorem{definition}[thm]{Definition}
\newtheorem{rem}[thm]{Remark}
\newtheorem{prf}[thm]{Proof}

\maketitle


\begin{abstract}
The Schr\"odinger cat states, constructed from Glauber coherent states and applied for description of qubits are  generalized to the kaleidoscope of coherent states, related with regular n-polygon symmetry and the roots of unity. This quantum kaleidoscope is motivated by our method of classical hydrodynamics images in a wedge domain, described by $q$-calculus of analytic functions with $q$ as a primitive root of unity.
The cases of the trinity states and the quartet states are described in details. Normalization formula for these states requires introduction of specific combinations of exponential functions with mod 3 and mod 4 symmetry.
We show that  these states can be generated for an arbitrary $n$ by the Quantum Fourier transform and can provide qutrits, ququats and in general, qudit  units of quantum information. Relations of our states with quantum groups and quantum calculus are discussed.
\end{abstract}

\section{Introduction}
\subsection{Classical Images and Vortex Kaleidoscope}
The problem of point vortices in a domain bounded by two circular cylinders can be reduced   to the Apollonius circles problem, corresponding to concentric circles \cite{p14}. Recently we have formulated the two circles theorem, allowing one to construct an arbitrary flow in such annular domain by the complex potential  as $q$-periodic analytic function, $F(q z) = F(z)$, where $q = R^2/r^2$ is determined by two circle radiuses. Depending on the number and the position of vortices, sources or sinks, we can fix singularity of this function in terms of $q$-elementary functions \cite{p16}.
  Similar theorem for the flow in the wedge domain  \cite{p15},  requires construction of $q^2$-periodic function for the complex potential $F(q^2 z) = F(z)$ and $q^2$- self-similar complex velocity $\bar V(q^2 z) = \bar q^2 \bar V(z)$, with $q$ as a root of unity $q^{2n} = 1$.
 This wedge theorem describes the fluid flow as superposition  of complex analytic functions
\begin{equation}
F(z) = \sum^{n-1}_{k=0} f(q^{2k} z) +  \sum^{n-1}_{k=0}\bar  f(q^{2k} z)\,, \label{qperiodicF}
\end{equation}
 representing the kaleidoscope of  images associated with the   regular $2n$ - polygon.  For the point vortex located at $z_0$, the theorem gives $q^2$-periodic complex potential
\begin{equation} \nonumber
F(z) = \frac{i \Gamma}{2 \pi} \ln \frac{z^n - z^n_0}{z^n - \bar z^n_0} = F(q^2 z),
\end{equation}
which due to the Kummer expansion
$$
z^n - z^n_0 = (z - z_0) (z - q^2 z_0) (z - q^4 z_0)... (z - q^{2(n-1)} z_0)\,,$$
 appears as the set of even  images at positions $z_0, q^2 z_0, q^4 z_0,..., q^{2(n-1)}z_0$ and the set of odd images at points $\bar z_0, q^2\bar  z_0, q^4\bar  z_0,..., q^{2(n-1)}\bar z_0$. This kaleidoscope of vortex images we called the Kummer kaleidoscope.
\subsection{Quantum Kaleidoscope and Coherent States}
Since analytical functions are related   intrinsically with quantum coherent states in the Fock-Bargman representation, here we extend our ideas to
 the Hilbert space for the coherent states. The problem is to construct $q$-periodic quantum states and $q$-self-similar quantum states. Such type of problem was studied recently  in \cite{Vitiello} as self-similarity properties of fractals  in the framework of the theory of entire analytical
functions and the q-deformed algebra of coherent states.
 In the present paper we consider the case, when $q$ is the root of unity $q^{2n} =1$ and show that it leads to the kaleidoscope of coherent states $| \alpha\rangle$,   $|q^2 \alpha\rangle$, ..., $|q^{2(n-1)} \alpha\rangle$,  located at vertices of the regular polygon.
 By acting with dilatation operator on analytic function
\begin{equation}
f(q^2 z) = q^{2 z \frac{d}{dz}} f(z)\label{dilatation}
\end{equation}
we can rewrite  the wedge theorem (\ref{qperiodicF}) in a compact form
\begin{equation}
F(z) = \sum^{n-1}_{k=0} \left( q^{2 z \frac{d}{dz}}\right)^k [f(z) + \bar f (z)] = [n]_{ q^{2 z \frac{d}{dz}}} [f(z) + \bar f (z)],\label{F}
\end{equation}
where we have used non-symmetric $q$-number with the  operator base. From this representation,
 $q^2$-periodicity for function $F(z)$ follows easily due to identity $ q^{2n z \frac{d}{dz}} = 1$.
We notice that the differential operator in (\ref{dilatation}) is the Fock-Bargman representation for the dilatation operator $q^{2 \widehat{N}}$, acting on coherent states as
\begin{equation} \nonumber
q^{2 \widehat{N}} | \alpha\rangle = |q^2 \alpha\rangle,
\end{equation}
where  $\widehat{N} = {a}^{\dagger} a$ is the number operator. Then, the  $q$-periodic coherent state analog of the wedge theorem (\ref{F}) is the quantum state
\begin{equation}\nonumber
 |0\rangle_{\alpha} \equiv (I + q^{2 \widehat{N}} +  q^{4\widehat{N}} + ... +  q^{2(n-1) \widehat{N}})  | \alpha\rangle =  [n]_{ q^{2 \widehat{N}}}  | \alpha\rangle.
\end{equation}
The $q$-periodicity for this quantum state $q^{2\widehat{N}} |0\rangle_{\alpha} = |0\rangle_{\alpha}$, follows easily from the relation $q^{2 n \widehat{N}} = I$.
This suggests also that to find  $q^2$- self-similar quantum  states, one can take the following superpositions of coherent states
$ [n]_{ q^{2 \widehat{N}+2}}  | \alpha\rangle, [n]_{ q^{2 \widehat{N}+4}}  | \alpha\rangle, ..., [n]_{ q^{2 \widehat{N}+ 2(n-1)}}  | \alpha\rangle .$ It turns out that this construction provides the set of orthogonal quantum states. The similar superpositions of coherent states were discussed in different context by several authors, as the generalized coherent states \cite{bb}, \cite{Stoler}, as factorization problem for the Schr\"odinger equation with self-similar potential \cite{Spiridonov} and as the Schr\"odinger cat states \cite{manko}.
The Schr\"odinger cat states as superposition of  Glauber's optical  coherent states with opposite phases, become important tool for construction of qubits, as a  units of quantum information in quantum optics. They correspond to even and odd quantum states with $q^2 = -1$.  Here we  generalize this construction  to the kaleidoscope of coherent states, related with regular n-polygon symmetry and the roots of unity. Superposition of coherent states with such symmetry plays the role of the quantum Fourier transform and provides the set of orthonormal quantum states,
 as a description of  qutrits, ququats and qudits. Such quantum states, considered  as a units of quantum information processing and corresponding to an arbitrary base number $n$, could have advantage in secure quantum communication.

\subsection{Glauber Coherent States}
We consider the Heisenberg-Weyl algebra,  written in terms of creation and annihilation operators, satisfying  bosonic commutation relation  $\left[\hat{a},\hat{a}^{\dag}\right]=\widehat{\textrm{I}}$.
The annihilation operator determines the vacuum state $ \hat{a}| 0 \rangle = 0 $ from the Hilbert space  $  | 0 \rangle \in \cal H $ and the
creation operator $ \hat{a}^{\dag}| 0 \rangle = | 1 \rangle$,  repeatedly applied to this state, gives  orthonormal set of states $  | n\rangle=\frac{( \hat{a}^{\dag} )^n}{\sqrt{n!}} | 0 \rangle$.

Coherent states are defined as eigenstates of annihilation operator : $  \hat{a}| \alpha \rangle = \alpha | \alpha \rangle ,  \alpha \in \mathcal{C} $. This gives us a relation between complex plane and the Hilbert space, such that $ \alpha \in \mathcal{C} \leftrightarrow | \alpha \rangle \in \cal H  $.
Another equivalent definition is given by  the displacement operator $D(\alpha)$,
\begin{equation} \nonumber
D(\alpha)=e^{ \alpha \hat{a}^{\dagger}-\bar{\alpha} \hat{a}}=e^{ -\frac{1}{2}|\alpha|^2 } e^{ \alpha \hat{a}^{\dagger} }  e^{-\bar{\alpha} \hat{a}} \,\,\,\Rightarrow \,\,\,
|\alpha \rangle=D(\alpha) |0\rangle  =e^{ -\frac{1}{2}|\alpha|^2 } \sum _{n=0}^{\infty} \frac{\alpha^n}{\sqrt{n!}} | n \rangle\,.
\end{equation}
By factorizing $D(\alpha)$, we get the following representation of  coherent states:
$|\alpha\rangle=\frac{e^{ \alpha \hat{a}^{\dagger}}}
{\sqrt{e^{|\alpha|^2}}}  |0\rangle   $, which is instructive for our next generalizations.
The inner product of coherent states, $$ \langle \alpha|\beta \rangle= e^{-\frac{1}{2}|\alpha|^2-\frac{1}{2}|\beta|^{2}+\bar{\alpha}\beta} \Rightarrow
  | \langle \alpha | \beta \rangle | ^2 = e^{-|\alpha - \beta|^2} $$ is never zero. This is why coherent states are not orthogonal states. Our aim is by using coherent states to construct an orthogonal set of states with discrete regular polygon symmetry.
\vspace{0.2 cm}

\section{Schr\"{o}dinger's Cat States}
\vspace{0.2 cm}
In description of the Schr\"{o}dinger cat states one introduces two orthogonal states as superpositions of $| \alpha \rangle$ and $|-\alpha \rangle$, which are called the even and the odd cat states,
$$| cat_{even}\rangle \equiv |\alpha_{+} \rangle \sim | \alpha \rangle + |-\alpha\rangle \quad , \quad |cat_{odd}\rangle \equiv |\alpha_{-} \rangle   \sim | \alpha \rangle - |-\alpha\rangle.$$
The states can be considered as a superposition of two coherent states related by rotation to angle $\pi$, which corresponds to primitive root of unity $q^2 =\overline{q}\,^2= -1$, so that $q^4=1$. The normalization constants  $N_{0},N_{1}$
are calculated as:
\begin{equation} \nonumber
 |0\rangle_{\alpha}= N_{0}(\, |\alpha \rangle + |q^2\alpha\rangle\, )\,,  \quad |1\rangle_{\alpha} = N_{1}(\,| \alpha \rangle + \overline{q}\,^2 |q^2\alpha\rangle\, )\,,
\end{equation}
\begin{equation}\nonumber
N_{0}=\frac{e^{\frac{|\alpha|^{2}}{2}}}{2\sqrt{\cosh|\alpha|^{2}}}\,,\quad
N_{1}=\frac{e^{\frac{|\alpha|^{2}}{2}}}{2\sqrt{\sinh|\alpha|^{2}}}\,.
\end{equation}
By acting with  the Hadamard gate,  these states can be represent in the matrix form:
\begin{equation}\nonumber
\left[
    \begin{array}{c}
      |0\rangle_{{\alpha}} \\
      |1\rangle_{{\alpha}} \\
    \end{array}
  \right] =\textbf{N}\underbrace{\frac{1}{\sqrt{2}}\left[\begin{array}{cc}
                                                1 & 1 \\
                                                  1 & \overline{q}\,^2 \\
                                                \end{array}
                                              \right]}_{Hadamard\,\,gate}\left[
                                                       \begin{array}{c}
                                                         | \alpha \rangle\\
                                                         | q^2\alpha \rangle \\
                                                       \end{array}
                                                     \right]\,,
                                                     \end{equation}
where the normalization matrix
\begin{equation} \nonumber
\textbf{N}=\frac{e^{\frac{|\alpha|^{2}}{2}}}{\sqrt{2}}\,\textrm{diag}\left( _{0}e^{|\alpha|^2}, \,_{1}e^{|\alpha|^2}\right)^{-1/2}(mod \,2)\equiv\textrm{diag}\left(N_0, N_1\right)
\end{equation}
is defined by the even ($0 \,\, mod \, 2$) and the odd ($1 \,\, mod\, 2$) exponential functions, coinciding with hyperbolic functions,
\begin{eqnarray} \nonumber
(mod \,2)\quad  _0 e^{|\alpha|^2}  &  \equiv &\sum _{k=0}^{\infty} \frac{\left(|\alpha|^{2}\right)^{2k}}{(2k)!}=\frac{e^{|\alpha|^{2}}+e^{{q}^{2}|\alpha|^{2}}}{2}
= \cosh |\alpha|^2, \\ \nonumber
(mod \,2) \quad _1 e^{|\alpha|^2}  & \equiv &\sum _{k=0}^{\infty} \frac{\left(|\alpha|^{2}\right)^{2k+1}}{(2k+1)!}=\frac{e^{|\alpha|^{2}}+ \bar q^2 e^{{q}^{2}|\alpha|^{2}}}{2} = \sinh |\alpha|^2.
\end{eqnarray}

\subsection{Different form of  Cat States}
In terms of these exponential functions
 we can rewrite the Schr\"{o}dinger cat states in a compact form:
\begin{eqnarray}\nonumber
|0\rangle_{{\alpha}}&=&\frac{_0 e^{\alpha\hat{a}^\dag} }{\sqrt{_0 e^{|\alpha|^2}}}|0\rangle \quad  (mod \,2) = \frac{\cosh \alpha \hat{a}^\dag}{\sqrt{\cosh |\alpha|^2}} |0\rangle\,, \\\nonumber
|1\rangle_{{\alpha}}&=&\frac{_1 e^{\alpha\hat{a}^\dag}}{\sqrt{_1 e^{|\alpha|^2}}}|0\rangle  \quad (mod \,2)= \frac{\sinh \alpha \hat{a}^\dag}{\sqrt{\sinh |\alpha|^2}} |0\rangle.
\end{eqnarray}
\subsection{Eigenvalue Problem for Cat States}
Since $|\alpha \rangle$ is an eigenstate of annihilation operator $\hat{a}$, $\hat{a}|\alpha \rangle = \alpha |\alpha \rangle $, it is also the eigenstate of operator $\hat{a}^2$:
$\hat{a}^2|\alpha\rangle=\alpha^2|\alpha\rangle$. However, the last equation admits one more eigenstate $ |-\alpha \rangle$ with the same eigenvalue $\alpha^2$, so that
$\hat{a}^2|\mp\alpha\rangle=\alpha^2|\mp\alpha\rangle$. Hence, any superposition of states $\{|+\alpha\rangle,|-\alpha\rangle\}$
 is also an eigenstate of operator $\hat{a}^2$, with the same eigenvalue.
This implies that Schr\"odinger cat states are eigenstates of this operator,
$$
\hat{a}^2| 0\rangle_{{\alpha}}=\alpha^2|0\rangle_{{\alpha}},
\hat{a}^2| 1\rangle_{{\alpha}}=\alpha^2|1\rangle_{{\alpha}},$$
 constituting  orthonormal basis $\{|0\rangle_{\alpha}, |1\rangle_{\alpha} \}.$
It can be used to define the qubit coherent state:
$$
| \psi \rangle_\alpha = c_0 | 0\rangle_{\alpha} + c_1 | 1\rangle_\alpha ,$$
where $|c_0|^2 + |c_1|^2 = 1$,
representing a unit of quantum information in quantum optics. This qubit state is an eigenstate of operator $\hat{a}^2$ as well:
$$
\hat{a}^2 | \psi \rangle_\alpha = \alpha^2 | \psi \rangle_\alpha.$$
\subsection{Number of Photons in Cat States}
\begin{figure}[h]
\begin{center}
\includegraphics[scale=0.60]{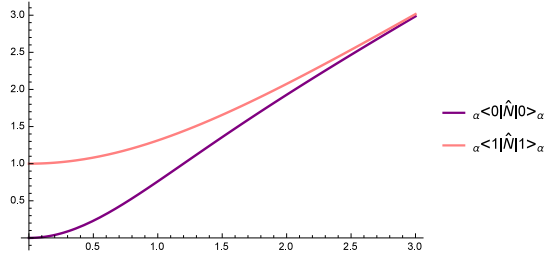}
\caption{Photon numbers in Schr\"{o}dinger's cat states} \label{fig1p}
\end{center}
\end{figure}

The annihilation operator $\hat{a}$ gives flipping between cat states $|0\rangle_{{\alpha}}$ and $|1\rangle_{{\alpha}}$:
\begin{eqnarray} \label{cp} \nonumber
\hat{a} |0\rangle_{{\alpha}}=\alpha\frac{N_0}{N_1}|1\rangle_{{\alpha}} \, , \quad \hat{a}|1\rangle_{{\alpha}}=\alpha\frac{N_1}{N_0}|0\rangle_{{\alpha}} \,.
\end{eqnarray}
By using these equations we find number of photons in  Schr\"{o}dinger's cat states as :
\begin{eqnarray} \nonumber
{_{\alpha}}\langle0|\widehat{N}|0\rangle_{{\alpha}}&=&|\alpha|^2\frac{N_0^2}{N_1^2}=|\alpha|^2\frac{_1 e^{|\alpha|^2}}{_0 e^{|\alpha|^2}} = |\alpha|^2 \tanh |\alpha|^2
 ,  \\ \nonumber
{_{\alpha}}\langle1|\widehat{N}|1\rangle_{{\alpha}}&=&|\alpha|^2\frac{N_1^2}{N_0^2}=|\alpha|^2\frac{_0 e^{|\alpha|^2}}{_1 e^{|\alpha|^2}}= |\alpha|^2 \coth |\alpha|^2,
\end{eqnarray}
(see Fig. $\ref{fig1p}$) and in the limiting case we have

$$\lim_{|\alpha|\to\infty}{_{\alpha}}\langle0|\widehat{N}|0\rangle_{{\alpha}}= \lim_{|\alpha|\to\infty}{_{\alpha}}\langle1|\widehat{N}|1\rangle_{{\alpha}}
\approx |\alpha|^2=\langle\pm\alpha|\widehat{N}|\pm\alpha\rangle.$$
\subsection{Schr\"{o}dinger's Kitten States : } In the limit
$$\lim_{|\alpha|\to0}{_{\alpha}}\langle0|\widehat{N}|0\rangle_{{\alpha}}=0 ,\quad
 \lim_{|\alpha|\to0}{_{\alpha}}\langle1|\widehat{N}|1\rangle_{{\alpha}}=1 , $$
we have so called Schr\"{o}dinger's kitten states with number of photons $0$ and $1$.
\subsection{Fermionic Representation of Cat States}

 The dilatation operator $q^{2\widehat{N}}= e^{i\pi \widehat{N}}=(-1)^{\widehat{N}}$ is the parity operator for cat states, so that
$|0\rangle_{{\alpha}}$ and $|1\rangle_{{\alpha}}$ states are eigenstates of this operator, the $q^2$-periodic one and $q^2$-self-similar one, correspondingly
\begin{equation} \nonumber
q^{2\widehat{N}} |0\rangle_{{\alpha}} = |0\rangle_{{\alpha}}\,,\,\,\,\,
q^{2\widehat{N}} |1\rangle_{{\alpha}} = q^2 |1\rangle_{{\alpha}}.
\end{equation}
These states represent  kaleidoscope of two coherent states, rotated by angle $\pi$,
\begin{equation} \nonumber
|0\rangle_{{\alpha}} = N_0 \,[2]_{q^{2\widehat{N}}} | \alpha \rangle = N_0 (I + q^{2\widehat{N}})| \alpha \rangle, \,\,\,|1\rangle_{{\alpha}} = N_1 \,[2]_{q^{2\widehat{N}+2}} | \alpha \rangle = N_1 (I +q^2 q^{2\widehat{N}})| \alpha \rangle
\end{equation}
or
\begin{equation} \nonumber
|0\rangle_{{\alpha}} = N_0 (I + (-1)^{\widehat{N}})| \alpha \rangle\,,\,\,\,|1\rangle_{{\alpha}} = N_1 (I - (-1)^{\widehat{N}})| \alpha \rangle\,.
\end{equation}
The cat states are eigenstates of $q^2$- number operator,
$$[\widehat{N}]{_{q^2}}=\frac{q^{2\widehat{N}}-1}{q^{2}-1} ,$$
where $q^2 = -1$, with eigenvalues $[0]_{q^2} = 0$ and $[1]_{q^2} = 1$.
In the Fock basis $|n \rangle $, $n=0,1,2...$, this number operator is diagonal,
with eigenvalues $[n]_{q^2} = 0$  for even numbers $n=2k$, and $[n]_{q^2} = 1$  for odd numbers $n=2k+1$.  The  matrix elements of number operator in cat basis are
$$
{_{\alpha}}\langle0|[\widehat{N}]{_{q^2}}|0\rangle_{{\alpha}}=[0]{_{q^2}}=0
, \quad
{_{\alpha}}\langle1|[\widehat{N}]{_{q^2}}|1\rangle_{{\alpha}}=[1]{_{q^2}}=1.
$$
\section{Trinity States}
Now, we consider the set of three coherent states rotated by angle $\frac{2\pi}{3}$ , which corresponds to $q^6=1$. First we define superposition $$|0\rangle_{\alpha}=\frac{N_{0}}{\sqrt{3}}\left(|\alpha\rangle+|q^{2}\alpha\rangle+|q^{4}\alpha\rangle \right).$$
Due to identity $$\, q^{6n}-1=(q^{2n}-1)(1+q^{2n}+q^{4n})=0 \, \Rightarrow \,
1+q^{2n}+q^{4n}=3\,\delta_{n \equiv 0(mod\,3)}, \qquad \label{mod3}
$$
we find normalization constant $N_{0}=e^{\frac{|\alpha|^{2}}{2}}(3\,_{0}e^{|\alpha|^2})^{-1/2},$
where we have introduced ($mod\, 3$) exponential function $$ \quad\displaystyle{_{0}e^{|\alpha|^2}
(mod \,3) \equiv \sum_{k=0}^{\infty}\frac{\left(|\alpha|^2\right)^{3k}}{(3k)!}= \frac{1}{3}\left( e^{|\alpha|^{2}}+e^{{q}^{2}|\alpha|^{2}}+e^{{q}^{4}|\alpha|^{2}} \right) }.$$
In a similar way we obtain the set of orthonormal states $|0\rangle_{\alpha},|1\rangle_{\alpha}$ and $ |2\rangle_{\alpha}$:
\begin{eqnarray}\nonumber
 |0\rangle_{\alpha}&=&e^{\frac{|\alpha|^{2}}{2}} \frac{| \alpha \rangle+|q\,^{2}\alpha\rangle+|q\,^{4}\alpha\rangle}
{\sqrt{3}\sqrt{e^{|\alpha|^{2}}+e^{{q}^{2}|\alpha|^{2}}+e^{{q}^{4}|\alpha|^{2}}}}
 =e^{\frac{|\alpha|^{2}}{2}} \frac{| \alpha \rangle + |q\,^{2}\alpha\rangle+|q\,^{4}\alpha\rangle }{3\sqrt{ _{0}e^{|\alpha|^2}{(mod \,3)}}} \, ,  \nonumber \\ \nonumber \\
 |1\rangle_{\alpha}&=& e^{\frac{|\alpha|^{2}}{2}} \frac{| \alpha \rangle+\overline{q}^{2}|q\,^{2}\alpha\rangle+\overline{q}^{4}|q\,^{4}\alpha\rangle}
{\sqrt{3}\sqrt{e^{|\alpha|^{2}}+\overline{q}^{2}e^{{q}^{2}|\alpha|^{2}}+\overline{q}^{4}e^{{q}^{4}|\alpha|^{2}}}}
= e^{\frac{|\alpha|^{2}}{2}} \frac{| \alpha \rangle +\overline{q}^{2} |q\,^{2}\alpha\rangle+
 \overline{q}^{4}|q\,^{4}\alpha\rangle }{3\sqrt{ _{1}e^{|\alpha|^2}(mod \,3)}}\, , \nonumber \\
|2\rangle_{\alpha}&=& e^{\frac{|\alpha|^{2}}{2}} \frac{| \alpha \rangle+\overline{q}^{4}|q\,^{2}\alpha\rangle+\overline{q}^{2}|q\,^{4}\alpha\rangle}
{\sqrt{3}\sqrt{e^{|\alpha|^{2}}+\overline{q}^{4}e^{{q}^{2}|\alpha|^{2}}+\overline{q}^{2}e^{{q}^{4}|\alpha|^{2}}}}
= e^{\frac{|\alpha|^{2}}{2}} \frac{| \alpha \rangle +\overline{q}^{4} |q\,^{2}\alpha\rangle+
 \overline{q}^{2}|q\,^{4}\alpha\rangle }{3\sqrt{ _{2}e^{|\alpha|^2}(mod \,3)}}\, .\nonumber
\end{eqnarray}
\subsection{Matrix form of Trinity States:}
The states appear by action of the trinity gate, playing role of three dimensional analogue of Hadamard gate
\begin{equation} \nonumber
\left[
  \begin{array}{c}
    |0\rangle_{\alpha} \\
    |1\rangle_{\alpha} \\
    |2\rangle_{\alpha} \\
  \end{array} \right]= \textbf{N}\underbrace{{\frac{1}{\sqrt{3}} \left[ \begin{array}{ccc}
                                                                     1 & 1 & 1 \\
                                                                     1 & \overline{q}^{2} & \left(\overline{q}^{2}\right)^{2} \\
                                                                     1 & \overline{q}^{4} & \left(\overline{q}^{4}\right)^{2} \\
                                                                   \end{array}
                                                                 \right]}}_{Trinity\,\,gate}\left[
                                                                          \begin{array}{c}
                                                                             | \alpha \rangle\\
                                                                             | {q}^{2}\alpha \rangle \\
                                                                             | {q}^{4}\alpha \rangle\\
                                                                          \end{array}
                                                                        \right] \,,
\end{equation}
with normalization constants
\begin{equation} \nonumber
  \textbf{N}=\frac{e^{\frac{|\alpha|^{2}}{2}}}{\sqrt{3}}\,\textrm{diag}\left( _{0}e^{|\alpha|^2}, _{1}e^{|\alpha|^2},_{2}e^{|\alpha|^2}\right)^{-1/2}(mod\,\, 3)\equiv\textrm{diag}\left(N_0, N_1, N_2 \right)
\end{equation} and identity
$$  1+\overline{q}^{2(n-k)}+\overline{q}^{4(n-k)}=3\,\delta_{n\equiv k(mod\,\,3)}\quad , 0\leq k \leq2 $$
\subsection{Phase Structure of Trinity States}
Trinity states as superposition of coherent states with explicit phase shift are following:
\begin{eqnarray} \nonumber
|0\rangle_{\alpha}&=& N_0 (| \alpha \rangle + |e^{i\frac{2\pi}{3}}\alpha\rangle+|e^{-i\frac{2\pi}{3}}\alpha\rangle ), \\ \nonumber
|1\rangle_{\alpha}&=& N_1 (| \alpha \rangle + e^{-i\frac{2\pi}{3}} |e^{i\frac{2\pi}{3}}\alpha\rangle+e^{i\frac{2\pi}{3}}|e^{-i\frac{2\pi}{3}}\alpha\rangle ), \\ \nonumber
|2\rangle_{\alpha}&=& N_2 (| \alpha \rangle + e^{i\frac{2\pi}{3}} |e^{i\frac{2\pi}{3}}\alpha\rangle+e^{-i\frac{2\pi}{3}}|e^{-i\frac{2\pi}{3}}\alpha\rangle ).
\end{eqnarray}

\subsection{Different form  of Trinity States}
By using three different $(mod\,\,3)$ exponential functions, we can rewrite our trinity states in a compact form:
\begin{eqnarray} \nonumber
|0\rangle_{{\alpha}}=\frac{_0 e^{\alpha\hat{a}^\dag} }{\sqrt{_0 e^{|\alpha|^2}}}|0\rangle\, , \quad
|1\rangle_{{\alpha}}=\frac{_1 e^{\alpha\hat{a}^\dag}}{\sqrt{_1 e^{|\alpha|^2}}}|0\rangle\, , \quad
|2\rangle_{{\alpha}}=\frac{_2 e^{\alpha\hat{a}^\dag}}{\sqrt{_2 e^{|\alpha|^2}}}|0\rangle \qquad (mod\,\,3).
\end{eqnarray}
\subsection{Eigenvalue Problem for Trinity States}
Coherent states $\{|\alpha\rangle,|q^2\alpha\rangle,|q^4\alpha\rangle \}$ are eigenstates of operator $\hat{a}$ with different eigenvalues $\alpha,q^2\alpha,q^4\alpha$, and the eigenstates of operator $\hat{a}^3$ with the same eigenvalue $\alpha^3$. Due to this, our trinity states $\{ |0\rangle_{\alpha}, |1\rangle_{\alpha},|2\rangle_{\alpha} \}$  are also eigenstates of operator $\hat{a}^3$  :
$$
\hat{a}^3|q^{2k}\alpha\rangle=\alpha^3|q^{2k}\alpha\rangle \quad \Rightarrow \quad
\hat{a}^3|k\rangle_{{\alpha}}=\alpha^3|k\rangle_{{\alpha}}  \quad k=0,1,2.
$$
From trinity states we can construct the qutrit coherent state
$
| \psi \rangle_\alpha = c_0 | 0\rangle_{\alpha} + c_1 | 1\rangle_\alpha + c_2 | 2\rangle_\alpha$, where
$|c_0|^2 + |c_1|^2 + |c_2|^2 =1$,
as a unit of quantum information with base $3$. It turns out that this state
is an eigenstate of operator $\hat{a}^3$:
$
\hat{a}^3 | \psi \rangle_\alpha = \alpha^3 | \psi \rangle_\alpha$.
\subsection{Number of Photons in Trinity States}
The annihilation operator $\hat{a}$ acts as flipping between states $|0\rangle_{{\alpha}},|1\rangle_{{\alpha}}$ and $|2\rangle_{{\alpha}}$:
\begin{eqnarray} \label{tp}
\hat{a}|0\rangle_{{\alpha}}=\alpha\frac{N_0}{N_2}|2\rangle_{{\alpha}}\,,\,\,
\hat{a}|1\rangle_{{\alpha}}=\alpha\frac{N_1}{N_0}|0\rangle_{{\alpha}} \, ,\,\,
\hat{a}|2\rangle_{{\alpha}}=\alpha\frac{N_2}{N_1}|1\rangle_{{\alpha}}  \, .
\end{eqnarray}
Equation $(\ref{tp})$ allows us to calculate number of photons in trinity  states (see Fig. 2):
\begin{eqnarray} \nonumber
{_{\alpha}}\langle0|\widehat{N}|0\rangle_{{\alpha}}&=&|\alpha|^2\left[\frac{_2 e^{|\alpha|^2}}{_0 e^{|\alpha|^2}}\right]=
|\alpha|^2\left[\frac{1+2e^{\frac{-3|\alpha|^2}{2}}\cos\left(\frac{\sqrt{3}}{2}|\alpha|^2+\frac{2\pi}{3}\right)}
{1+2e^{\frac{-3|\alpha|^2}{2}}\cos\left(\frac{\sqrt{3}}{2}|\alpha|^2\right)}\right]
 \,, \\ \nonumber
{_{\alpha}}\langle1|\widehat{N}|1\rangle_{{\alpha}}&=&|\alpha|^2\left[\frac{_0 e^{|\alpha|^2}}{_1 e^{|\alpha|^2}}\right]=
|\alpha|^2\left[\frac{1+2e^{\frac{-3|\alpha|^2}{2}}\cos\left(\frac{\sqrt{3}}{2}|\alpha|^2\right)}
{1+2e^{\frac{-3|\alpha|^2}{2}}\cos\left(\frac{\sqrt{3}}{2}|\alpha|^2-\frac{2\pi}{3}\right)} \right]
 \, ,\\ \nonumber
{_{\alpha}}\langle2|\widehat{N}|2\rangle_{{\alpha}}&=&|\alpha|^2\left[\frac{_1 e^{|\alpha|^2}}{_2e^{|\alpha|^2}}\right]=
|\alpha|^2\left[\frac{1+2e^{\frac{-3|\alpha|^2}{2}}\cos\left(\frac{\sqrt{3}}{2}|\alpha|^2-\frac{2\pi}{3}\right)}
{1+2e^{\frac{-3|\alpha|^2}{2}}\cos\left(\frac{\sqrt{3}}{2}|\alpha|^2+\frac{2\pi}{3}\right)} \right]\,.
\end{eqnarray}

\begin{figure}[h]
\begin{center}
\includegraphics[scale=0.60]{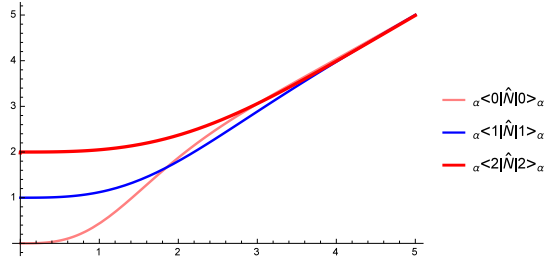}
\caption{Photon numbers in Trinity States} \label{fig2p}
\end{center}
\end{figure}
Since $\widehat{N}| n\rangle=n|n\rangle, n\geq0 $ , for the $q^2$-number operator  $[\widehat{N}]{_{q^2}}=\frac{q^{2\widehat{N}}-1}{q^{2}-1}$,  due to
$$
q^{2\widehat{N}}|0\rangle_{{\alpha}}=|0\rangle_{{\alpha}} , \quad
q^{2\widehat{N}}|1\rangle_{{\alpha}}=q^2|1\rangle_{{\alpha}} , \quad
q^{2\widehat{N}}|2\rangle_{{\alpha}}=q^4|2\rangle_{{\alpha}}\,,
$$
we have diagonal form with matrix elements
$$
{_{\alpha}}\langle0|[\widehat{N}]{_{q^2}}|0\rangle_{{\alpha}}=[0]{_{q^2}}, \quad
{_{\alpha}}\langle1|[\widehat{N}]{_{q^2}}|1\rangle_{{\alpha}}=[1]{_{q^2}}, \quad
{_{\alpha}}\langle2|[\widehat{N}]{_{q^2}}|2\rangle_{{\alpha}}=[2]{_{q^2}}\,.
$$
\subsubsection{Matrix Representation}
 We find matrix representation of the operators in our kaleidoscope basis as the clock and the shift matrix
\begin{equation} \nonumber
q^{2\widehat{N}} = \left(
                    \begin{array}{ccc}
1 & 0 & 0  \\
0 & q^2 & 0  \\
0 & 0 & q^4\end{array}
                           \right),\,\,\,
   \hat{a} = \alpha \left(
                    \begin{array}{ccc}
0 & \frac{N_1}{N_0} & 0  \\
0 & 0 & \frac{N_2}{N_1}  \\
\frac{N_0}{N_2} & 0 & 0\end{array}
                           \right)   =  \alpha \left(
                    \begin{array}{ccc}
\frac{N_1}{N_0} & 0 & 0  \\
0 & \frac{N_2}{N_1} & 0  \\
0 & 0 & \frac{N_0}{N_2}\end{array}
                           \right) \left(
                    \begin{array}{ccc}
0 & 1 & 0  \\
0 & 0 & 1  \\
1 & 0 & 0\end{array}
                           \right)\,.
\end{equation}

\section{Quartet States}
We defined four states,  rotated by  angle $\frac{\pi}{2}$ and determined by primitive root of unity: $q^8 =1$.
Superposition of these states with proper coefficients give us quartet of orthonormal basis states:
\begin{equation} \nonumber
  \left[
    \begin{array}{c}
      |0\rangle_{\alpha} \\
      |1\rangle_{\alpha} \\
      |2\rangle_{\alpha} \\
      |3\rangle_{\alpha}\\
    \end{array}
  \right]=\textbf{N}\underbrace{\small{ \frac{1}{\sqrt{4}}\left[\begin{array}{cccc}
                                                      1 & 1 & 1 & 1 \\
                                                      1 & \overline{q}^{2} & \left(\overline{q}^{2}\right)^{2} &\left(\overline{q}^{2}\right)^{3} \\
                                                      1 & \overline{q}^{4} & \left(\overline{q}^{4}\right)^{2} & \left(\overline{q}^{4}\right)^{3} \\
                                                      1 & \overline{q}^{6}& \left(\overline{q}^{6}\right)^{2} & \left(\overline{q}^{6}\right)^{3} \\
                                                    \end{array}
                                                  \right]}}_{Quartet\,\,gate}\left[
                                                           \begin{array}{c}
                                                             | \alpha \rangle \\
                                                             | {q}^{2}\alpha \rangle \\
                                                             | {q}^{4}\alpha \rangle \\
                                                             | {q}^{6}\alpha \rangle \\
                                                           \end{array} \right] \, ,
\end{equation}
where normalization constants are defined as $$\,\, \displaystyle{ \textbf{N}=\frac{e^{\frac{|\alpha|^{2}}{2}}}{\sqrt{4}}\,\textrm{diag}\left( _{0}e^{|\alpha|^2}, _{1}e^{|\alpha|^2},_{2}e^{|\alpha|^2},_{3}e^{|\alpha|^2}\right)^{-1/2}(mod\, 4)\equiv\textrm{diag}\left(N_0, N_1, N_2, N_3\right) }$$
and the identity is
$$ 1+\overline{q}^{2(n-k)}+\overline{q}^{4(n-k)}+\overline{q}^{6(n-k)}=4\,\delta_{n\equiv k(mod \,4)} \quad ,0\leq k \leq3 $$
\subsection{Phase Structure of Quartet States}
The quartet states are superpositions of cat states with explicit form of phase shift as
\begin{eqnarray} \nonumber
|0\rangle_{\alpha}&=& N_0 \left[(| \alpha \rangle + |-\alpha\rangle) +(|i\alpha\rangle +|-i\alpha\rangle) \right] , \\ \nonumber
|1\rangle_{\alpha}&=& N_1 \left[(| \alpha \rangle - |-\alpha\rangle) -i (|i\alpha\rangle -|-i\alpha\rangle) \right] , \\ \nonumber
|2\rangle_{\alpha}&=& N_2 \left[(| \alpha \rangle + |-\alpha\rangle) - (|i\alpha\rangle +|-i\alpha\rangle) \right] ,\\  \nonumber
|3\rangle_{\alpha}&=& N_2 \left[(| \alpha \rangle - |-\alpha\rangle) +i (|i\alpha\rangle -|-i\alpha\rangle) \right].
\end{eqnarray}

\subsection{Different form of  Quartet States}

By using $(mod \, 4)$ exponential functions we get representation of quartet states in a compact form:
$$
|0\rangle_{{\alpha}}=\frac{_{0} e^{\alpha\hat{a}^\dag} }{\sqrt{_0 e^{|\alpha|^2}}}|0\rangle, \quad
|1\rangle_{{\alpha}}=\frac{_{1} e^{\alpha\hat{a}^\dag}}{\sqrt{_1 e^{|\alpha|^2}}}|0\rangle , \quad
|2\rangle_{{\alpha}}=\frac{_{2} e^{\alpha\hat{a}^\dag}}{\sqrt{_2 e^{|\alpha|^2}}}|0\rangle  , \quad
|3\rangle_{{\alpha}}=\frac{_{3} e^{\alpha\hat{a}^\dag}}{\sqrt{_3 e^{|\alpha|^2}}}|0\rangle \,\, (mod\,4).
$$

\subsection{Eigenvalue Problem for Quartet States}
As easy to see the quartet states are eigenstates of operator $\hat{a}^4$ with eigenvalue $\alpha^4$:
$$
\hat{a}^4|q^{2k}\alpha\rangle=\alpha^4|q^{2k}\alpha\rangle \quad \Rightarrow \quad
\hat{a}^4|k\rangle_{{\alpha}}=\alpha^4|k\rangle_{{\alpha}}  \quad k=0,1,2,3.
$$
\subsection{Number of Photons in Quartet States}
The annihilation operator $\hat{a}$ implements  flipping between states $|k\rangle_{{\alpha}} ,\,\, k=0,1,2,3$ :
\begin{eqnarray} \label{qp}
\hat{a}|0\rangle_{{\alpha}}=\alpha\frac{N_0}{N_3}|3\rangle_{{\alpha}} \, ,\,\,
\hat{a}|1\rangle_{{\alpha}}=\alpha\frac{N_1}{N_0}|0\rangle_{{\alpha}}  \,,\,\,
\hat{a}|2\rangle_{{\alpha}}=\alpha\frac{N_2}{N_1}|1\rangle_{{\alpha}} \,,\,\,
\hat{a}|3\rangle_{{\alpha}}=\alpha\frac{N_3}{N_2}|2\rangle_{{\alpha}} \,\,.
\end{eqnarray}
By using $(\ref{qp})$ we calculate number of photons in quartet states (See Fig. 3):
\begin{eqnarray} \nonumber
{_{\alpha}}\langle0|\widehat{N}|0\rangle_{{\alpha}}&=&|\alpha|^2\left[\frac{_3 e^{|\alpha|^2}}{_0 e^{|\alpha|^2}}\right]
=|\alpha|^2\left[\frac{\sinh|\alpha|^2-\sin|\alpha|^2}{\cosh|\alpha|^2+\cos|\alpha|^2}\right] \,,\\ \nonumber
{_{\alpha}}\langle1|\widehat{N}|1\rangle_{{\alpha}}&=&|\alpha|^2\left[\frac{_0 e^{|\alpha|^2}}{_1 e^{|\alpha|^2}}\right]
=|\alpha|^2\left[\frac{\cosh|\alpha|^2+\cos|\alpha|^2}{\sinh|\alpha|^2+\sin|\alpha|^2}\right] \,,\\ \nonumber
{_{\alpha}}\langle2|\widehat{N}|2\rangle_{{\alpha}}&=&|\alpha|^2\left[\frac{_1 e^{|\alpha|^2}}{_2 e^{|\alpha|^2}}\right]
=|\alpha|^2\left[\frac{\sinh|\alpha|^2+\sin|\alpha|^2}{\cosh|\alpha|^2-\cos|\alpha|^2}\right] \,,\\ \nonumber
{_{\alpha}}\langle3|\widehat{N}|3\rangle_{{\alpha}}&=&|\alpha|^2\left[\frac{_2 e^{|\alpha|^2}}{_3 e^{|\alpha|^2}}\right]
=|\alpha|^2\left[\frac{\cosh|\alpha|^2-\cos|\alpha|^2}{\sinh|\alpha|^2-\sin|\alpha|^2}\right]\,. \nonumber
\end{eqnarray}

\begin{figure}[h]
\begin{center}
\includegraphics[scale=0.60]{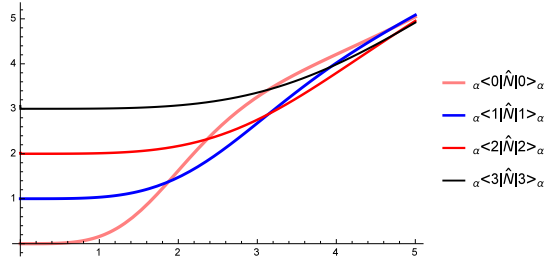}
\caption{Photon numbers in quartet states} \label{fig3p}
\end{center}
\end{figure}
The quartet states are eigenstates of $q^2$-number operator $[\widehat{N}]{_{q^2}}$:
\begin{equation} \nonumber
q^{2\widehat{N}}|k\rangle_{{\alpha}}=q^{2k}|k\rangle_{{\alpha}}\quad \Rightarrow \quad [\widehat{N}]{_{q^2}}|k\rangle_{{\alpha}}=[k]{_{q^2}}|k\rangle_{{\alpha}} \,\, , \textrm{where}\,\, k=0,1,2,3.
\end{equation}
In terms of these states we can describe the ququats quantum state as a unit of quantum information with base 4.

\section{Kaleidoscope of Quantum Coherent States}
As a generalization of previous results here we consider superposition of $n$ coherent states, which are belonging to vertices of regular $n$-polygon and  rotated by angle $\frac{\pi}{n}$ (Fig. \ref{fig3}).  It is related with primitive root of unity: $q^{2n}=1$. For the inner product of $q^{2k}$ rotated coherent states we have
$\langle q^{2k}\alpha |q^{2k}\alpha \rangle =1$, $\langle q^{2k}\alpha |q^{2l}\alpha \rangle =e^{|\alpha|^2({q}^{2(l-k)}-1)}   ,\, 0\leq k,l \leq n-1 $.\\

Lemma: For $q^{2n}=1 \, , \, 0\leq s \leq n-1$
\begin{itemize}
  \item $1+q^{2m}+q^{4m}+...+q^{2m(n-1)}= n\delta_{m\equiv0(mod\, n)}$
  \item $1+q^{2(m-s)}+q^{4(m-s)}+...+q^{2(m-s)(n-1)}= n\delta_{m\equiv s(mod \,n)}$
\end{itemize}

\subsection{Quantum Fourier Transformation}

\begin{figure}[h]
\begin{minipage}{14pc}
\includegraphics[scale=0.21]{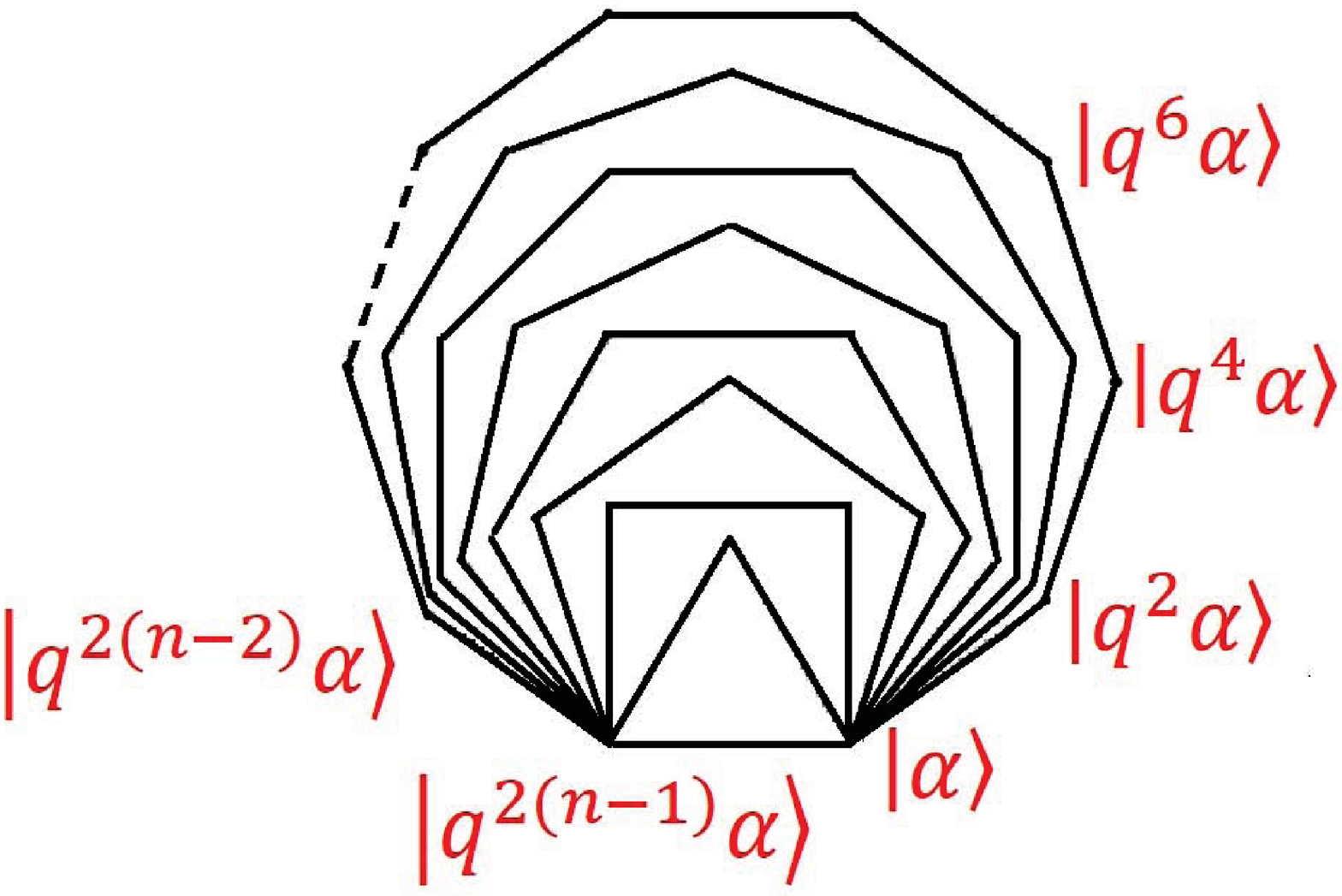}
\caption{The regular n-polygon} \label{fig3}
\end{minipage}
\hspace{4pc}%
\begin{minipage}{14pc}
\begin{equation} \nonumber
\left[
   \begin{array}{c}
     |0\rangle_{\alpha} \\
     |1\rangle_{\alpha} \\
     |2\rangle_{\alpha} \\
     |3\rangle_{\alpha} \\
     \vdots \\
   \tiny{ |n-1\rangle_{\alpha} } \\
   \end{array}
 \right]=\textbf{NQ} \left[
   \begin{array}{c}
     |\alpha\rangle \\
     |q^2\alpha\rangle \\
     |q^4\alpha\rangle \\
     |q^6\alpha\rangle \\
     \vdots \\
    \tiny{ |q^{2(n-1)}\alpha\rangle }\\
   \end{array}
 \right]
\end{equation}
\caption{General structure of kaleidoscope states} \label{fig4}
\end{minipage}
\end{figure}

Our construction (Fig. \ref{fig4}) shows that orthogonal kaleidoscope states can by described by the Quantum Fourier transform
\begin{equation}
\left[
   \begin{array}{c}
     |\widetilde{0}\rangle_{\alpha} \\
     |\widetilde{1}\rangle_{\alpha} \\
     |\widetilde{2}\rangle_{\alpha} \\
     |\widetilde{3}\rangle_{\alpha} \\
     \vdots \\
   \tiny{ |\widetilde{n-1}\rangle_{\alpha} } \\
   \end{array}
 \right]=\frac{1}{\sqrt{n}}\left[
                    \begin{array}{cccccccc}
1 & 1 & 1 & ... & 1 \\
1 & w & w^{2} & ... & w^{n-1}  \\
1 & w^{2} & w^{4} & ... & w^{2(n-1)}\\
1 & w^{3} & w^{6} & ... & w^{3(n-1)}\\
\vdots & \vdots & \vdots & \ddots & \vdots \\
1 & w^{(n-1)}  & w^{2(n-1)} & ...  & w^{(n-1)(n-1)} \\
                             \end{array}
                           \right] \left[
   \begin{array}{c}
     |\alpha\rangle \\
     |q^2\alpha\rangle \\
     |q^4\alpha\rangle \\
     |q^6\alpha\rangle \\
     \vdots \\
    \tiny{ |q^{2(n-1)}\alpha\rangle }\\
   \end{array}
 \right],
\end{equation}
where $w=e^\frac{-2\pi i}{\,n} = \bar q^2$ is the $n$-th root of unity, so that corresponding unitary matrix $QQ^{\dag}=Q^{\dag}Q=I$ satisfies:

\begin{equation}\nonumber
\boxed{|\widetilde{k}\rangle_{\alpha}\longmapsto\frac{1}{\sqrt{n}}\sum_{j=0}^{n-1}w^{jk}|q^{2j}\alpha\rangle \quad 0\leq k \leq n-1}
\end{equation}
For orthonormal states we define normalization matrix,
$$\,\, \displaystyle{ \textbf{N}=\frac{e^{\frac{|\alpha|^{2}}{2}}}{\sqrt{n}}\,\textrm{diag}\left( _{0}e^{|\alpha|^2}, _{1}e^{|\alpha|^2},_{2}e^{|\alpha|^2},...,_{n-1}e^{|\alpha|^2}\right)^{-1/2}(mod\, n) }$$
in terms of $(mod\,\,n)$ exponential functions:
\begin{equation}
 f_{s}(|\alpha|^2)=_{s}e^{|\alpha|^2}(mod\, n)\equiv \sum_{k=0}^{\infty}\frac{(|\alpha|^2)^{nk+s}}{(nk+s)!}\,\,,\quad 0\leq s \leq n-1\,. \label{f}
\end{equation}
These functions represent superposition of standard exponentials
\begin{equation} \nonumber
_{s}e^{|\alpha|^2}(mod\,n)=\frac{1}{n}\sum_{k=0}^{n-1} \overline{q}^{2sk}e^{{q}^{2k}|\alpha|^{2}}\,\,,\quad 0\leq s \leq n-1\,,
\end{equation}
related to each other by derivatives
$$
\frac{\partial}{\partial |\alpha|^{2}} \left[ {}_{s}e^{|\alpha|^2}\right] = {}_{s-1}e^{|\alpha|^2} \,\, , \quad \frac{\partial}{\partial |\alpha|^{2}} \left[_{0}e^{|\alpha|^2}\right]= _{n-1}e^{|\alpha|^2} \, .
$$
According to this, function $f_{s}$ defined in ($\ref{f}$) is a solution of ordinary differential equation
\begin{equation}
 f^{(n)}_{s}= f_{s} \, ,\quad\textrm{where} \,\,  0\leq s \leq n-1 , \label{difeqn}
\end{equation}
with proper initial values: $ f^{(s)}_s (0) = 1$ and
$f_s(0) = f'_s(0) = ...= f^{(s-1)}_s (0) = f^{(s+1)}_s (0) = ...= f^{(n-1)}_s (0) = 0$.
\subsection{Representation of Kaleidoscope of Quantum Coherent States:}

\begin{equation}|\alpha \rangle=e^{ -\frac{1}{2}|\alpha|^2 } e^{ \alpha \hat{a}^{\dagger} }| 0 \rangle \, \Rightarrow \,
|s \rangle_{\alpha}=\frac{_{s}e^{ \alpha \hat{a}^{\dagger}}}
{\sqrt{_{s}e^{|\alpha|^2}}}| 0 \rangle \quad (mod \, n), \,\,\,0\leq s \leq n-1. \label{nkaleidoscope}\end{equation}

\subsection{Number of Photons in Kaleidoscope of Quantum Coherent States:}
\begin{equation} \nonumber
\hat{a}|s\rangle_{{\alpha}}=\alpha\frac{N_{s}}{N_{s-1}}|s-1\rangle_{{\alpha}}\quad \Rightarrow\quad
{_{\alpha}}\langle s|\widehat{N}|s \rangle_{{\alpha}}=|\alpha|^2\left[\frac{_{s-1}e^{|\alpha|^2}}{_{s}e^{|\alpha|^2}}\right]  \,\, , 1<s \leq n-1\,,
\end{equation}
\begin{equation} \nonumber
\hat{a}|0\rangle_{{\alpha}}=\alpha\frac{N_{0}}{N_{n-1}}|n-1\rangle_{{\alpha}}\quad\Rightarrow \quad
{_{\alpha}}\langle 0|\widehat{N}|0 \rangle_{{\alpha}}= |\alpha|^2\left[\frac{_{n-1}e^{|\alpha|^2}}{_{0}e^{|\alpha|^2}}\right]\,,
\end{equation}
$$\lim_{|\alpha|\to\infty}{_{\alpha}}\langle s|\widehat{N}|s\rangle_{{\alpha}}
 \approx |\alpha|^2=\langle\pm\alpha|\widehat{N}|\pm\alpha\rangle\,,\,\,\,\,\lim_{|\alpha|\to0}{_{\alpha}}\langle s|\widehat{N}|s\rangle_{{\alpha}}=s\,.$$

\section{Quantum Algebra}
Our kaleidoscope coherent states (\ref{nkaleidoscope}) are eigenstates of operator $q^{2\hat N}$: $q^{2\hat N} |k \rangle_{\alpha} = q^{2k}|k \rangle_{\alpha}$, $k=0, 1...,n-1$. In the Fock space this operator is an infinite matrix of the form
\begin{equation}
\Sigma_3 \equiv q^{2\hat N} = I \otimes \left(
                    \begin{array}{cccccccc}
1 & 0 & 0 & ... & 0 \\
0 & q^2 & 0 & ... & 0  \\
0 & 0 & q^4 & ... & 0\\
\vdots & \vdots & \vdots & \ddots & \vdots \\
0 & 0  & 0 & ...  & q^{2(n-1)} \\
                             \end{array}
                           \right),\,\,\,
\Sigma_1 = I\otimes \left(
                    \begin{array}{cccccccc}
0 & 0 & 0 & ... & 1 \\
1 & 0 & 0 & ... & 0  \\
0 & 1 & 0 & ... & 0\\
\vdots & \vdots & \vdots & \ddots & \vdots \\
0 & 0  & 0 & ...  & 0 \\
                             \end{array}
                           \right)\,.
\end{equation}
Here the $n \times n$ matrices are called the Sylvester clock and shift matrices correspondingly. They are $q$-commutative
$\Sigma_1 \Sigma_3 = q^2 \Sigma_3 \Sigma_1$, satisfy relations $\Sigma^n_1 = I$, $\Sigma^n_3 = I$ and
are connected  by the unitary transformation:
$\Sigma_1 = (I \otimes Q) q^{2\hat N} (I \otimes Q^+)$. Hermann Weyl in book \cite{weyl}  proposed  them for description of quantum mechanics of finite dimensional systems.

By dilatation operator $q^{2\hat N}$ we define $q^2$-number operator $[\hat N]_{q^2} = (q^{2\hat N} - 1)/(q^2 -1)$ for non-symmetrical $q$-calculus, and $[\hat N]_{\tilde q^2} = (q^{2\hat N} - q^{-2\hat N})/(q^2 -q^{-2})$ for the symmetrical one. In our kaleidoscope basis, these number operators are diagonal and given by q-numbers: $[\hat N] = diag ([0], [1],..., [n-1]) $. For symmetric case the q-number operator is Hermitian and can be factorized as
$ [\hat N] = \hat B^+ \hat B$, $[
\hat N+1] = \hat B \hat B^+$, where $\hat B = \hat a \sqrt{\frac{[N]_{\tilde q^2}}{N}}$ explicitly is
\begin{equation}
\hat B = I\otimes  \left(
                    \begin{array}{cccccccc}
0 & \sqrt{[1]} & 0 & ... & 0 \\
0 & 0 & \sqrt{[2]} & ... & 0  \\
\vdots & \vdots & \vdots & \ddots & \vdots \\
0 & 0  & 0 & ...  & 0 \\
                             \end{array}
                           \right),\,\,\,
                           \hat B^+ = I \otimes \left(
                    \begin{array}{cccccccc}
0 & 0 & 0 & ... & 0 \\
\sqrt{[1]} & 0 & 0 & ... & 0  \\
0 & \sqrt{[2]} & 0 & ... & 0  \\
\vdots & \vdots & \vdots & \ddots & \vdots \\
0 & 0  & 0 & ...  & 0 \\
                             \end{array}
                           \right)
\end{equation}
and $\hat B^n = 0$, $(\hat B^+)^n = 0$. In non-symmetric case the number operator is not Hermitian.

\subsection{Symmetric Case} For symmetric case we have the quantum algebra
\begin{equation}
\hat B \hat B^+ - q^2 \hat B^+ \hat B = q^{-2\hat N},\,\,\, \hat B \hat B^+ - q^{-2}\hat B^+ \hat B = q^{2\hat N}\,,
\end{equation}
and  quantum $q^2$-oscillator with Hamiltonian
$
\hat H = \frac{\hbar \omega}{2}\left( [\hat N]_{\tilde q^2} + [\hat N+ I]_{\tilde q^2}\right)$.
On the  kaleidoscope states as eigenstates, the spectrum of this Hamiltonian is
\begin{equation}
E_k = \frac{\hbar \omega}{2}\frac{\sin \frac{2\pi}{n} (k+ \frac{1}{2})}{\sin \frac{\pi}{n}}.\nonumber
\end{equation}
The same spectrum was obtained in \cite{floratos} for description of two anyons system. Appearance of quantum algebraic structure in two different physical systems, as optical coherent states and the anyons problem is instructive.

\subsection{Non-symmetric Case} Algebra of these operators is $q^2$-deformed
\begin{equation}
\hat B \hat B^+ - q^2 \hat B^+ \hat B = I,\,\,\, \hat B \hat B^+ - \hat B^+ \hat B = q^{2\hat N}\,,\nonumber
\end{equation}
 with periodic (mod n) ($[k+n]_{q^2} = [k]_{q^2}$) $q^2$-numbers
 \begin{equation}
 [k]_{q^2} = e^{i\frac{\pi}{n}(k-1)} \frac{\sin \frac{\pi}{n} k}{\sin \frac{\pi}{n}}\,.\nonumber
 \end{equation}

Kaleidoscope of coherent states considered in present paper can be realized by proper phase superposition of coherent states of light (the Gaussian states) and it can provide a unit of quantum information corresponding not only to diadic, but also to an arbitrary  number base $n$. These states furnish the representation of quantum symmetry related with  quantum q-oscillator. As a generalization of the Schr\"odinger cat states, from our kaleidoscope states one can construct multi qudits entangled quantum states. This work is in progress.

\section{Acknowledgements} This work is supported by TUBITAK grant 116F206.

\end{document}